\documentclass[pre,aps,onecolumn,showpacs,floatfix,superscriptaddress]{revtex4-1}
\pdfoutput=1
\usepackage{graphicx}
\usepackage{dcolumn}
\usepackage{color}
\usepackage{hyperref}
\usepackage{latexsym}
\usepackage{amsmath, amsthm, amssymb}
\usepackage{epsfig}
\usepackage{latexsym}
\usepackage{bm}
\usepackage{enumitem}
\begin{document}
\title{Achieving perfect coordination amongst agents in the co-action minority game}
\author{ Hardik Rajpal}
\affiliation{\small{Centre for Complexity Science and Department of Mathematics, Imperial College London, South Kensington Campus, London SW7 2AZ, UK}}
\author{Deepak Dhar}
\affiliation{\small{Department of Physics, Indian Institute for Science Education and Research,Pune 411008, India}}

\date{\today}

\begin{abstract}
We discuss the strategy that  rational agents can use  to maximize their expected long-term payoff in the co-action minority game. We argue that the agents will try to get into a cyclic state, where each of the $(2N +1)$ agent wins exactly $N$ times in any continuous stretch of $(2N+1)$ days. We propose and analyse a strategy for reaching such a cyclic state quickly, when any direct communication between agents is not allowed, and only the publicly available common information is the record of total number of people choosing the first restaurant in the past. We determine exactly the average time required to reach the periodic state for this strategy. We show that it varies as $(N/\ln 2) [1 +  \alpha \cos (2 \pi \log_2 N)$], for large $N$, where the amplitude $\alpha$ of the leading term in the log-periodic oscillations is found be $\frac{8 \pi^2}{(\ln 2)^2} \exp{(- 2 \pi^2/\ln 2)} \approx {\color{blue}7 \times 10^{-11}}$.

\end{abstract}

\maketitle
\section{Introduction}

Since its introduction by Challet and Zhang in 1997, the~Minority Game (MG)  has become a~
prototypical model for the~study of behavior of a~system of interacting agents~\cite{challet-zhang}. In the~original model, the~key element is the~so-called ``bounded rationality'' of the~agents: the~agents are adaptive, and have limited information about the~behavior of other agents. They use simple rules to~decide their next move, based on the~past outcomes.  It is found that, for a~range of parameter values, the~system of agents evolves into~a~ state, where the~global efficiency is larger than possible if they simply made a~random choice independent of history of the~game.  This is one of the~simplest models of  learning and adaptation by agents, that shows the~complex emergent phenomena~of  self-organization, and coevolution in a~system of  many interacting agents, and has attracted much interest. The~ model can be solved exactly in the~limit of large number of agents and large times, although this requires  rather sophisticated mathematical techniques, e.g., functional integrals, and the~replica~method. There are nice reviews of the~existing work on MG~\cite{moro, kets}, and monographs~\cite{challet-marsili-zhang, coolen}.

 However, to~understand the~effect of bounded rationality of the~agents, it is imperative to~compare this with the~case where the~rules of the~game are same as before, but agents are rational. This led  Sasidevan and Dhar to~introduce a~variation of the~standard MG, called the~Co-action Minority game (CAMG) in which the~ allowed moves, the~stipulation about no direct communication between agents, and the~payoffs for different outcomes are kept unchanged, but  the~agents are assumed to~be fully rational, and their choice of possible strategies is not constrained in any way ~\cite{sasidevan-dhar}.  Studying the~difference of its steady state from that of the~standard Challet--Zhang Minority Game (CZMG) underscores  the~role of bounded rationality in the~latter.

In the~CAMG, the~agents can use mixed strategies, and try to~optimize their payoff not for the~next day, but a~weighted sum of their future payoffs, with higher weights for more immediate gains.\mbox{ It was found that} the~optimal strategy depends on the~time horizon of agents, and  overall efficiency of the~system increases if the~agents have a~larger time horizon.

 In this context, a~natural question arises: what would rational agents do if they aim to~maximize the~expectation value of their {\it long-term gain}. In this paper, we show that rational agents  can actually achieve the~maximum possible payoff, by getting into~a~cyclic state, and discuss how they can arrive at the~coordination needed for such a~state. In contrast, in CZMG, where agents have only bounded rationality,  the~expected payoff per agent is lesser than than this by an amount of order $a/\sqrt{N}$, same as if they chose restaurants randomly, but with a~smaller coefficient $a$.  We show that the~average time to~reach the~desired cyclic state satisfies a~functional equation. We solve this equation exactly and show that the~time required to~reach the~stationary state varies linearly with $N$.

 We note that agents can only use the~limited information publicly available to~achieve this coordination. The~problem becomes a~coordination game~\cite{crawford}, in the~setting of an MG. In~\cite{sasidevan-dhar}, it was assumed that agents decide their action based  only on the~previous day's result. Here, we drop this assumption, and assume that agents have access to~entire history of attendance record, and can also remember the~entire history of their own earlier actions. 

A~problem equivalent to~the~one studied in this paper is this: we have $(2N+1)$ agents playing a~minority game, where $N$ is a~positive integer. What is the~strategy the~agents can use  to~work together to~assign a~unique identification  number (ID) from $1 $ to~$2N +1$ to~each agent, such that  each agent knows her own ID,  in least time, using only the~public information of attendances in the~past?  Even more generally, we can think of agents that cannot communicate directly with each other, and each agent can only communicate with some central authority. The~authority  can send  messages to~agents,  only in the~broadcast mode, where the~same message is sent to~all the~agents. 

Even with this constraint, it is clearly possible for the~agents to~get unique IDs. For example, a~simple strategy would be that each agent first generates a~random string of  some length $m$, and sends it to~the~central authority. We choose $m$ to~be large enough ~that the probability of two different agents generating the~same string is small ( say, $m \approx 2 \log N$). Then, the~central authority arranges these bits in some ordered list (of total length  $mN$ bits), and broadcasts the~list. Then, the~agent can infer his ID from the~the~position of her unique string in the~list. However, this scheme is clearly not optimal. 
What is the~least number of bits that have to~be broadcast  to~assign unique IDs to~all agents, so that each agent knows her own ID?  In our problem, the~agents need to~coordinate using only the~information of the~past attendance record, and their own past actions.

We describe a~particular strategy to~achieve this coordination. This strategy is quite efficient, in that the~coordination is achieved in a~time that increases only linearly with the~number of agents, but we have no proof that  it is the~best possible. 
  Interestingly, we find that the~mean time to~reach the~cyclic  shows  log-periodic oscillations.  In addition, the~amplitude of the~oscillations is very small (of order $10^{-11}$). Such a~small value, {\it obtained without any fine-tuning},  seems quite unexpected. The~mathematical mechanism  involved  may be of interest in the~more general context of understanding how  many natural systems select some  very small parameter-values (e.g., the~inverse correlation length in self-organized critical systems, or the~cosmological constant~\cite{cosmological}). It is also of interest in the~analysis of algorithms, where these kinds of oscillations were first encountered, and studied~\cite{log-periodic1, log-periodic2, odlyzko}.

This paper is organized as follows. In Section~\ref{section2}, we define the~model precisely. In Section~\ref{section3}, we  discuss  the~complication due to~possibility of coalition formation in the~game, but argue that rational agents who, by definition, optimize their personal long-time average payoff, will aim to~reach  a~periodic state. In Section~\ref{section4}, we describe a~coordination strategy that will reach the~cyclic state, using only the~publicly available attendance information. In Section~\ref{section5}, we study the~average time required to~reach the~periodic state, as a~function of the~number of agents. We show that the~generating function for the~average times satisfies a~functional equation in one variable. We solve this equation  exactly, and  find that the~expected time, when the~number of agents is $2N+1$,  asymptotically increases as  $N /(\ln 2)$ for large $N$, and shows log-periodic oscillations, but of a~very small amplitude of order $10^{-11}$. In the~final Section~\ref{section6}, we summarize our results, and mention another model of resource allocation, called the~Kolkata~Paise Restaurant problem, where the~optimal state is also periodic. We also discuss the~relation of  this study to~other problems  showing log-periodic oscillations.

\section{Definition of the~Model}
\label{section2}
We consider a~set of $(2 N + 1)$ agents, who choose between two options (say choosing one of the~two restaurants A~or B for dinner) every day. The~assumption of total number of players being odd  is a~simplifying assumption, standard in MG literature,  as then we need not  specify additional rules about the~payoffs in case of a~tie. Every day,  agents in the~restaurant with smaller attendance (i.e., less than or equal to~$N$) get a~payoff of $1$, while the~rest get nothing. In choosing which restaurant to~go to, the~agents can not communicate with each other directly, and the~only information available to~them is the~number of agents $(2N+1)$, and the~entire history of the~number of people who chose a~in the~past. This public information is naturally the~same for  all agents. In addition, an agent can remember her own history of choices in the~past, which constitutes her private information.  

In the~original formulation of the~MG, as defined by Challet and Zhang, (hereafter referred to~as the~CZMG), the~agents are adaptive, and try to~maximize their expected payoff, {\it for the~next day}.  Each agent chooses the~restaurant, based on one of the~strategies from a~small set of strategies available to~him. In the~CZMG, the~agents assign performance-based scores to~the~strategies available to~them, and use the~one with the~highest score.  While this is  perhaps a~reasonable  first model of the~behavior of agents in some  real world situations, it is not particularly efficient~\cite{sornette}, and we would like to~explore other possible strategies of agents, to~see if they can perform better.   

 In CAMG, the~agents are rational, and are allowed to~use mixed strategies, and decide the~weights of different options  {\it{rationally}} themselves. In addition, unlike CZMG, where each agent tries to~maximize her expected payoff the {\it {next day}}, here the~agents optimize the~average discounted payoff per day $\bar{W}$, defined~as
\begin{equation}
\bar{W} =(1 - \alpha) \sum_{r=1}^{\infty} \alpha^r W_r,
\end{equation}
where $W_r$ is the~expected payoff on the~r-th future day, and $\alpha~<1$ is called the~discount parameter.   In~\cite{sasidevan-dhar}, it is shown that the~choice of optimal strategy by agents depends on $\alpha$, and changes  discontinuously  as $\alpha$ is varied continuously. In particular, in this paper, we consider the~special case where $\alpha$ tends to~$1$, which corresponds to~the~the~limit where each agent tries to~maximize  the~long-time average of her  expected payoff per day. 

\section{Optimality of the~Cyclic State} 
\label{section3}
In the~Minority Game, by definition, the~maximum number of winning agents on any particular day is less than or equal to~$N$. Thus, if the~expected payoff per day, averaged over all agents, is $\bar{P}$, we have the~obvious inequality
\begin{equation}
\bar{P} \leq \frac{N}{2 N +1}.
\label{ineq:1}
\end{equation}

It is easy to~construct a~situation where this inequality is saturated: Consider the~case where each agent visits the~restaurant A~for $N$ consecutive days, and then goes to~B for the~next $(N+1)$ days, and they coordinate  their periodic schedules such that on any particular day, there are exactly $N$ people in the~restaurant A~(this is clearly possible). Then, in 
such a~state, each person's time-averaged expected payoff per day is $N/(2 N +1)$.

From the~symmetry between the~agents in the~definition of the~model, all agents start with same information, and have the~same time horizon.  Then, clearly,  the~expected long-time average payoff per day  will be the~same for each agent. However, in the~rules of the~Minority Game defined above, there is a~possibility of coalition formation, where, if  some agents successfully reach an understanding, called here a~coalition,  then they may
achieve an average payoff greater than $N/(2 N +1)$, while other agents, not part of the~coalition, receive an average payoff strictly less than $N/(2 N +1)$. 

This may be seen most easily when $N=1$. Here, there are exactly three agents,  called X, Y and~Z. Then, X and Y may reach an agreement that X uses a~periodic pattern, say  BAABA, and Y uses a~complimentary pattern ABBAB. Then, whatever choice Z makes, he will always be in the~majority. Then, his average payoff is zero, and the~combined payoff of X and Y is $1$ per day. If Z makes his choice at random, the~average payoff of X (or, equivalently Y) is $1/2 > \bar{P}= 1/3$.

How can X and Y reach such an understanding, without any direct communication? In general, this may happen by accident. For example, if agents are choosing at random, X and Y may notice that, in the~recent past, they win more often if they choose the~specific periodic patterns. Then, X~and Y have reason to~stay with these choices, and they have managed to~form a~coalition, without any direct communication. We note that the~coalition is formed, without  the~partners  knowing  each others'~identities!

If somehow, in our game, X and Y manage to~form a~coalition, clearly, Z is at a~disadvantage. If~this happens, Z  could try to~retaliate by choosing a~periodic string of same period. Then, his payoff remains zero, but  it is possible that the~ payoff of X or Y becomes less than $1/3$.  Clearly, then, it would become disadvantageous for that agent to~stay in the~coalition. Unfortunately, Z has no way to~infer this period from the~available information, and  can only make a~guess, and see if it works.  

A~selfish agent X will prefer to~get into~a~coalition, as then his expected payoff would be greater  than $1/3$. However, he cannot be sure to~form such  a~coalition, and there is a~non-zero probability of him being the~person outside the~coalition formed with zero payoff. By symmetry between the~agents, this probability is $1/3$. 

Under these circumstances, would an agent prefer to~look for an uncertain coalition, where he may be punished, or the~partner could defect anytime, or would he prefer a~coordinated equitable cyclic state where everyone gets an  guaranteed average payoff  of $1/3$? Clearly, one cannot reach any conclusion about the~psychological preferences of agents from the~definition of the~model given so far.  This requires a~further specification. In the~following, we will assume  that {\it {rational agents, by definition, want to} maximize their expected pay-off, and hence will  prefer the~cyclic state with a~certainty of getting a~payoff of $N$ every $(2N +1)$ days, to~the~uncertain coalition state. } We note that, in a~cyclic state, any single agent has no incentive to~deviate from the~common strategy, if all others are following it.

We note that the~ higher payoffs  possible in the~coalition state for an agent are offset by the~higher probability of doing worse. From the~inequality in Equation (\ref{ineq:1}), any  other strategy can, at best, equal the~average long-term payoff  obtained in a~cyclic state. Thus, rational agents  will  prefer the~cyclic state over others.

\section{A~Coordination Strategy to~Reach the~Periodic State}
\label{section4}
As explained above, rational agents  will prefer to~get into~a~cyclic state. The~simplest cyclic state is of period $(2 N +1)$.  Of course, there are many cyclic states possible, and the~strategy should enable the~agents to~coordinate their behavior to~strive towards  the~same cyclic state.  In addition, to~maximize their expected payoffs, the~agents will like to~reach this cyclic state, in as short a~time as possible. 

 To~reach this coordination, all agents {\it have to} follow some common strategy. The~existence of a~common ``common-sense'' strategy that all agents follow does not contradict the~assumption of no prior or direct communication between agents in the~game.  This may be seen most simply in a~much simpler coordination problem: consider the~``Full-house Game'', where the~payoff is $1$ for all, only if all people are in the~same restaurant. Else, everybody gets $0$. Then, there is no conflict of interests between agents, but complete coordination between their actions is still required.
For this trivial game, there is an obvious  common-sense strategy: on the~first day, people choose the~restaurant at random.  Then, on second day,  every one goes to~the~restaurant that had more people, and stays with the~same choice for all subsequent days. Then, after the~first day, every agent wins every day. However, note that if all agents do not follow the~strategy, it will not work.

In fact, clearly, no strategy for coordination can work, if all agents do not follow it. Thus, the~question is: if the~agents can  infer the~strategy, will others  follow? Only if they can do this consistently,  coordination can be achieved.

 As is well-known, in choosing among equal states, otherwise extraneous considerations can become important for reaching a~consensus. For example, in choosing land boundary between two countries, one may pick some geographical feature such as a~ridge, or a~river. For the~ example of Full-house Game above, another possible, equally effective, strategy is that  all agents switch their choice, every day after the~second day. Which of these two should be adopted by the~agents?  In this case, the~agents can reasonably, and consistently, argue that the~first strategy is ``simpler'', and hence more likely to~be selected by all. 

These same considerations apply to~the~strategy we  discuss below. In trying to~decide what strategy other agents will follow, we argue that rational agents will assume that all agents will choose a~strategy that is most efficient, and, if there are several strategies that pass this criterion, they will choose the~simplest out of them.  We  try to~convince the~reader that our proposed strategy meets these criteria, and hence would be selected by  future players of the~game.  We  give  an example below of  a~strategy that would also work, but is less efficient, and less simple. While we can not prove that our proposal is the~simplest possible, it is the~simplest amongst the~ones we could think of, and no other strategy is known to~be more efficient.

Let us first describe the~overall structure of the~strategy.  On any day, the~agent knows the~history of attendances so far, and based on this either decides to~stay with the~same choice as previous day, or changes her choice with a~shift probability, using a~personal random number generator. The~shift probability is specified by the~strategy, and depends on the~history of the~game so far. The~strategy consists of two stages. We start  at day $t =0$ with $(2N+1)$ agents.   At the~end of the~ the~first stage, the~agents have divided themselves into~ two groups: the~first consisting of exactly $N$ agents, and the~second having the~remaining $(N+1)$ agents,  and each agent knows to~which group she belongs. In this algorithm, agents in the~first group assign  the~same  ID  $0$ to~themselves. In the~second stage, the~remaining agents, by their coordinated actions, distribute unique IDs, labeled $1$ to~$N+1$ amongst themselves.  At the~end of the~second stage, each agent knows his own ID, and also knows when all assignments are complete.   

After this, setting up a~cyclic state is straight forward, but still requires coordination. We will make a~specific choice of this cycle.   This particular choice seems  natural (in the~sense discussed above),  and is as follows: all agents with ID $0$, on all days, choose option A. 
An agent with non-zero ID $r$  will choose option B on all days, except the~day  $t =   2 r-1 [mod (2N +1)]$, when she chooses option A.   

Clearly, this produces a~cyclic state with period $(2 N +1)$.  Let the~days be marked cyclically as $1, 2, 3  \ldots (2N+1)$, starting with the~day after  assignments of IDs is complete. The~agents with ID $0$ are  the~winning minority  on Days $2, 4, 6\ldots$~On any odd day, exactly $N$ out of the~$(N+1)$ agents with non-zero ID  have choice B, and are the~winning minority, and the~person left out is different on different odd days.  

This specific choice is also assumed to~known to, and selected by  all agents.  In our case, simplicity of the~algorithm to~get there, and quickness in reaching the~desired state are guiding criteria~that lead to~ this choice.  What distinguishes this particular cyclic state from others possible are two special features: one is that, on any given day, exactly one, or at most two, persons shift their choice of the~restaurant, and this minimizes the~number of moves made. This choice  has another distinguishing feature:  not only each agent wins on exactly $N$ days, out of $(2 N +1)$ days. In fact, any agents wins at least $m -1 $ times, and at most $m$ times, in any consecutive period of $ 2 m$ days, for all positive integers~$m$. We do not have a~  proof that this particular state is uniquely selected by these criteria, but it is the~best among several we could think of. 

Now, we specify the~algorithm in the~first stage. At the~start of the~algorithm, on day $t=0$, all~agents are in the~same state. We may say that all agents were in the~same restaurant on the~
previous day. Then, each agents chooses the~restaurant to~go to~the~next day randomly:  the~agent generates a~random number from her personal random number generator, and chooses option A~iff the~random number is~$\leq$1/2, and else chooses option B. 
At the~end of Day 1, it is announced publicly how many people went to~A, and how many to~B. Let the~number of people going to~the~minority restaurant be $ (N - \Delta)$, and then, clearly, the~remaining $(N +1 +\Delta)$ went to~the~majority restaurant. 

If $\Delta~=0$, the~first stage ends. Else, for $\Delta~\neq 0$, the~people in the~minority restaurant stay back in the~same restaurant, but each agent in the~majority restaurant shift with a~probability $\Delta/(N + \Delta~+1)$. Then, on next day, there will be some more people moving into~the~minority restaurant. Let the~number of people in the~new minority restaurant be $N -\Delta'$. This number $\Delta'$ is a~random variable. If it takes the~value $0$, the~algorithm ends. Else, the~same procedure is repeated. The~agents iterate the~procedure until they achieve an exact $(N, N+1)$ split between the~restaurants. The~ $N$ agents 
in the~minority restaurant are assigned the~ID $0$, and stay in the~same restaurant for all subsequent times, until the~completion of the~ID assignments, and, later, in the~cyclic state.

Now, we specify the~strategy in the~second stage. This is also a~recursive algorithm. 
Suppose at any one call of the~algorithm, some, $R$ agents have to~be assigned IDs from a~list of $R$  items. On~ the~first day, these agents jump at random, while others stay with the~same choice as before. Thus, this group breaks into~two approximately equal subgroups, say of sizes $r_1$ and $r_2$, with $r_1 +r_2 =R$. Then, the~algorithm recursively assigns  to~the~smaller set of $r_1$ agents the~first $r_1$  items from the~list, and then the~remaining $r_2$  items to~the~second set.

Now, we specify the~strategy in more detail. 
Suppose at any one call of the~algorithm, some, say $N + R$ agents, have been already assigned IDs: $N$ agents have ID $0$, and $R$ of them have been assigned IDs from $1$ to~$R$, and  there is an identified set of  $r$  agents who are to~be assigned  the~next available set of IDs, from $R+1$ to~$R+r$. All these agents are in the~same restaurant, and know that they will now be assigned IDs, and the~set of IDs to~be assigned is also public knowledge. The~remaining $(N +1 -R)$ agents also know that they will not to~be assigned IDs in this call of algorithm (have already been assigned, or have to~wait further). This group remains with their current choice until all the~r agents have been assigned IDs. At the~beginning of the~execution of the~algorithm, $R=0, r = N+1$, and the~available IDs are $1$ to~$(N+1)$. 

The~algorithm for the~second stage is defined recursively as follows:\\ 
\vspace{-18pt}
	
\begin{itemize}[leftmargin=*,labelsep=6mm]
\item If $r=1$, clearly, the~only agent is assigned the~available ID, and he knows his ID, and the~algorithm~ends.\\
\vspace{-12pt}
\item If $r > 1$, the~agents that are not to~be assigned IDs in this round continue with the~same choice as previous day, until the~end of the~algorithm. The~agents use their personal random number generators to~break this set of $r$ agents into~two smaller roughly equal sets, of sizes $j_1$ and $r-j_1$, where $j_1$ is a~random variable.  Now, the~algorithm recursively assigns to~the~first set the~IDs from $R+1$ to~$R+j_1$, and then the~remaining IDs from $R  +j_1 + 1$ to~$R+r$ to~the~second set, and the~algorithm~ends.
\end{itemize}

The~break-up of the~$r$-set into~two parts is achieved as follows: each of the~ $r$ agents flips a~coin, and shift to~the~other restaurant with probability $1/2$. The~number of agents that actually shift is a~random variable, call it $j$. From the~attendance record next day, the~value of $j$ becomes known to~all agents. Now, the~set of $r$ to-be-assigned agents has been divided into~two smaller sets: one consisting of $j$ agents, in one restaurant, and the~remaining of $r-j$ agents, in the~other restaurant. If $j \neq r/2$, the~smaller set is called the~first set, and the~larger set the~second set, and we put $j_1 = Min(j, r-j)$. If $ j =r/2$, both sets are equal size, and then, the~set of agents that shifted is called the~first set, and the~others the~second set.

As all agents monitor the~attendance record, on any day, each agent knows how many IDs have been assigned so far, to~which set she belongs,  at what stage the~algorithm is, and what she should do next day (stay with her existing choice, or shift with probability $1/2$). In addition, every agent knows her own ID at the~end of the~algorithm.

\begin{figure}
\centering
\includegraphics[width=11.0cm]{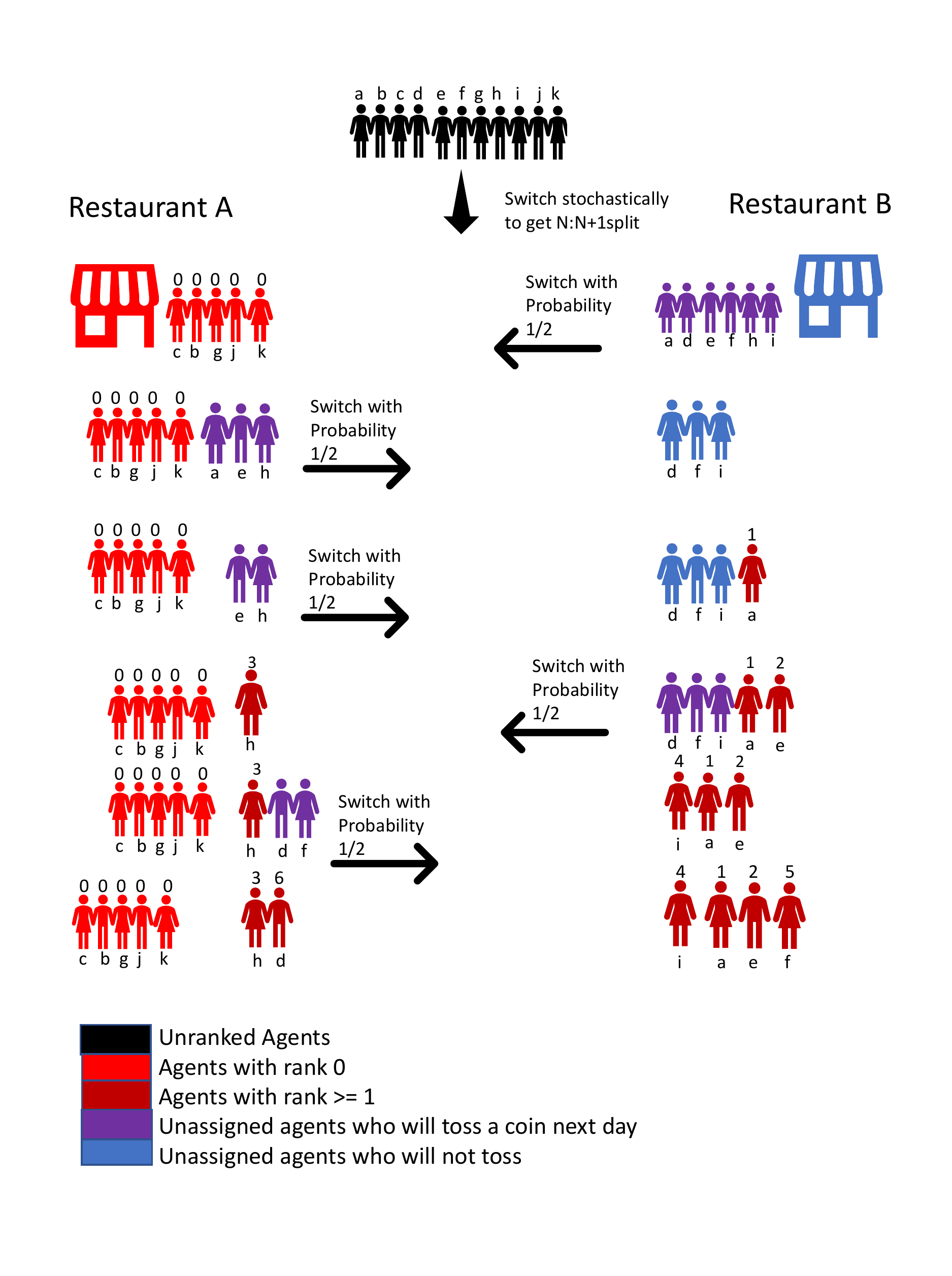}
\caption{Ranking of $N$ = 11 agents using the~proposed algorithm. The~numerals above the~agents indicates their assigned IDs. } 
\label{figure1}
\end{figure}

 Let us consider an example for the~case of say, $2N+1 = 11$ agents in Figure~\ref{figure1}. The~agents are identified by the~lower case letters $a, b, c, d ...k$.  For example, at the~end of the~first stage, the~agents  have achieved the~5:6 split. Suppose~agents $b, c, g, j, k$ are in ~restaurant A~and the~rest in B. Then, the~agents in A~are assigned ID $0$, and know it. On the~second day, Each of $a, d, e, f, h, i$ tosses a~coin to~decide to~switch  to~the~other restaurant or not. For example,  $a, e, h$ get heads, and actually switch. At the~end day,  all~agents know that three agents actually switched, and hence $a, e, h$ will be assigned  one of the~IDs from the~set $\{ 1,2,3\}$, and $d, f, i$ from the~set $\{4, 5, 6\}$. Then, on Day 3, $a, e$ and $ h$ toss a~coin. For example, $a$ gets head, and actually switches.  Then, at the~end of the~day, it is known that $a$ is assigned the~ID $1$. The~next day, only $e$  and $h$ flip a~coin, and rest stay with the~previous days choice. If $e$ actually jumps, then, at the~end of day, everybody knows that IDs $2$ and $3$ have been assigned, and $e$ knows that his is $2$, and $h$ knows that her ID  is 3. 
  The next day, $d, f$ and $i$ toss coins, and the rest stay with their previous day's choice, and so on.

\section{Expected Time to~Reach the~Cyclic State}
\label{section5}
Firstly, we argue that the~time required to~complete the~first stage, and assign rank zero to~exactly $N$ agents is of order $\log \log N$. When the~agents jump with probability $1/2$ on the~first day, the~expected number of agents in the~first restaurant is $(2 N +1)/2$, and standard deviation of this number is ${\cal O}(\sqrt{N})$. Let us say that the~number of people in the~majority restaurant on any given day is $(N + \Delta~+1)$. As the~probability of jump is $\Delta/(N + \Delta~+1)$, with $N \gg \Delta$, the~actual number of people jumping is a~random variable, the~distribution being approximately Poisson, with mean $\Delta$, and standard deviation approximately $\Delta^{1/2}$. Thus, we see that, on each day, the~logarithm of deviation decreases to~approximately half its value on previous day, until it becomes ${\cal O}(1)$, and, after that, the~expected time to~reach $N:(N+1)$ split is finite. Thus, the~total time to~complete the~first stage is $\log \log N$, and may be neglected, for large $N$~\cite{sasidevan-dhar-bkc}.

Now, we consider the~time required for the~second stage of the~algorithm. Let the~average time required to~assign unique IDs in the~second stage of the algorithm to~a~set of $n$ agents be $T_n$. Then, clearly, we have 
\begin{equation}
T_0 = T_1 =0.
\end{equation}
 
We now show that $T_2 =2$. Note that, at the~start of algorithm, both agents are in the~same restaurant. Then, with probability $1/2$, when both jump, they will end up in different restaurants, and the~assignment is done. Else, with probability $1/2$, both agents are in the~same restaurant, and the~state is the same as before. Hence, we must have
\begin{equation}
T_2 = 1 + (1/2) T_2,
\end{equation}
which implies that $T_2 =2$.  

This argument is easily extended to~higher values $n$. At the~start, all $n$ agents are in the~same restaurant. Then, as each agent chooses to~shift with probability $1/2$, the~probability that exactly $r$ people shift is $ \binom{n}{r}  2^{-n}$. However, then, the~ expected time for completion is $T_r + T_{n-r}$. Taking into~account the~one extra~day, we see that $T_n$ satisfies the~linear equation
\begin{equation}
T_n = 1 + \sum_{r=0}^{n} {\rm Prob}(r) \left[ T_r + T_{n-r} \right], {\rm ~~for ~ n \geq 2}.
\end{equation}

Since $ {\rm Prob}(r) =  {\rm Prob}(n - r)$, this equation may be simplified to
\begin{equation}
T_n = 1 + 2 \sum_{r=0}^{n}  {\rm Prob}(r)  T_r , {\rm ~~for ~ n \geq 2}.
\end{equation}

\begin{figure}
\centering
\includegraphics[width=11.0cm]{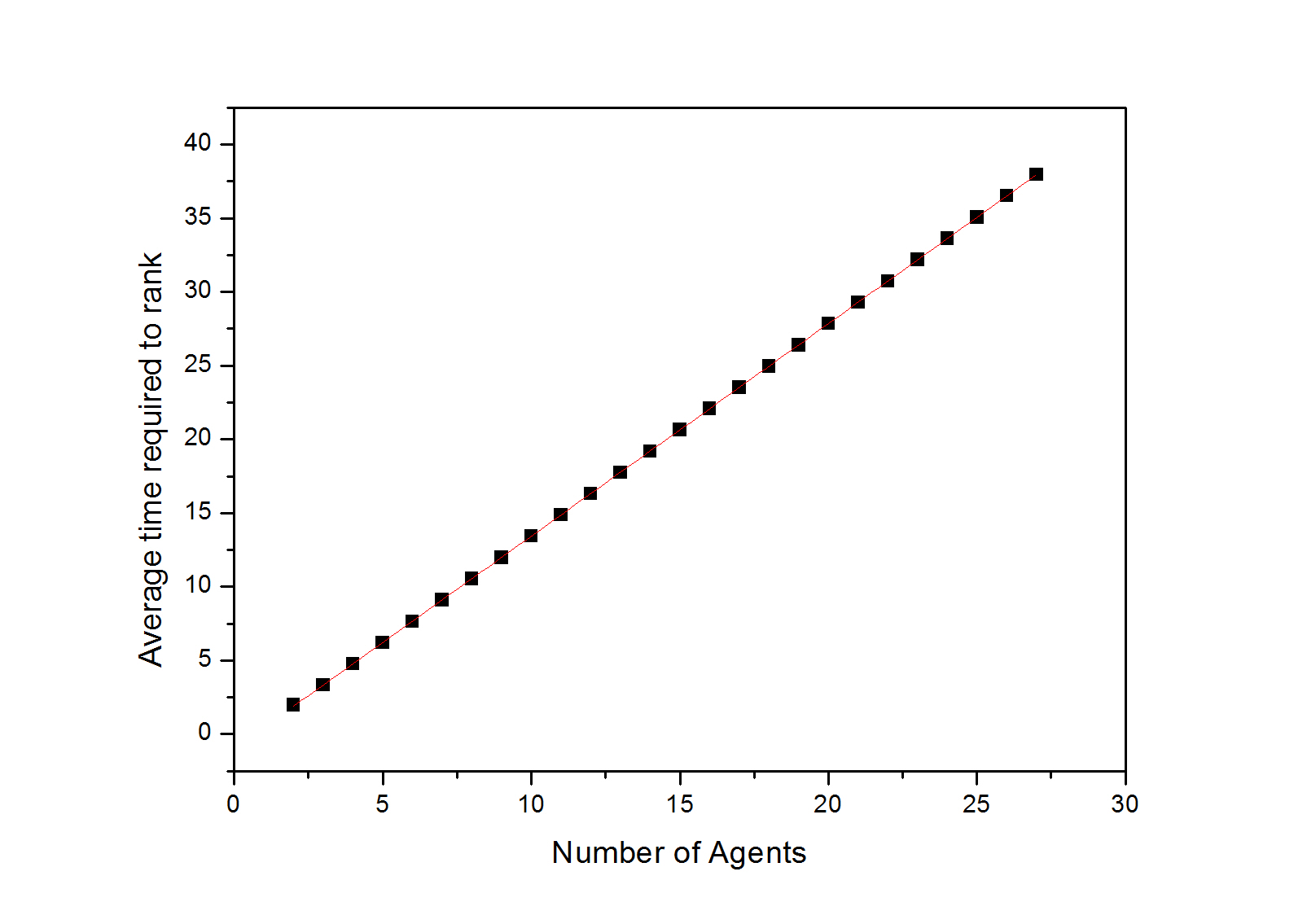}
\caption{Numerically determined exact values of $T_n$ for $n \leq 30$. The~equation of the~approximate linear fit here is $y = 1.4449x - 1.0451$.  }
\label{figure2}
\end{figure}

This is a~linear equation that relates $T_n$ to~values of $T_r$, with $r < n$. We can thus determine all the~$T_n$ recursively. For example, we get $T_3 = 10/3$, and $T_4 = 100/21$. The~resulting values of $T_n$, for $ n \leq 30$ were determined numerically, using a~simple computer program, and are shown in Figure~\ref{figure2}. We see that $T_n$ increases approximately as $1.4449 n$.

Define the~generating function 
\begin{equation}
T(x) = \sum_{r=1}^{\infty} T_r x^r,
\end{equation}
we see that $T(x)$ satisfies the~equation
\begin{equation}
T(x) = \frac{x^2}{(1-x)} + \sum_{n=2}^{\infty} \sum_{r=0}^{n}  {\rm Prob}(r) T_r x^n.
\label{eq:t1}
\end{equation}

We write $n = r+s$, then the~summation 
over  $s$  runs from $0$ to~$+\infty$, independent of the~value of $r$, and, noting that  ${\rm Prob}(r) = \binom{n}{r} 2^{-n}$, this can be done explicitly

\begin{equation}
\sum_{s=0}^{\infty} \binom{r+s}{r} x^s 2^{-s} = (1 - x/2)^{-r -1}
\end{equation}

Then,~Equation (\ref{eq:t1}) becomes
\begin{equation}
T(x) = \frac{x^2}{(1 - x/2)} + \frac{4}{(2-x)} T(\frac{x}{2-x})
\label{eq:t2}
\end{equation}
\unskip

As a~check, for small $x$, $T(x)$ is approximately $2 x^2$, which is consistent with above. For $x$ near $1$, we get $ T(x = 1 -\epsilon)  \approx 4 T(x = 1 - 2 \epsilon)$, which implies that $T(1 - \epsilon)$ diverges as $\epsilon^{-2}$, for small $\epsilon$. This then implies that $T_n$ varies linearly with $n$, for large $n$. To~be more precise, we can find finite constants $K_1$ and $K_2$ such that $K_1 n \leq T_n \leq K_2 n$. 

We can simplify Equation (\ref{eq:t2}) by making a~change of variables $x = 1/(y+1)$. Then, in terms of the~new variable $y$, we write  
\begin{equation}
T(x = \frac{1}{1+y}) = \frac{1+y}{y^2} \tilde{H}(y). 
\end{equation}
the~equation for $\tilde{H}(y)$ is seen to~be
\begin{equation}
\tilde{H}(y) = \frac{y}{(1+y)^2} +\tilde{H}(2y)
\end{equation}

This equation is easily solved by iteration, giving 
\begin{equation}
\tilde{H}(y) = \sum_{s=0}^{\infty} \frac{y 2^s}{(2^s y +1)^2}.
\label{eq:tildeH}
\end{equation}

The~asymptotic behavior of this function for $y$ near $0$  determines the~behavior of $H(x)$, for $x$ near $1$.  For very small $y$, we can extend the~lower limit of summation in Equation (\ref{eq:tildeH}) to~$-\infty$. Then, the~function $\tilde{H}(y)$ tends to~ $H^*(y)$, which is defined by: 
\begin{equation}
H^*(y) = \sum_{s = -\infty}^{+\infty}  \frac{y 2^s}{(2^s y +1)^2}.
\label{H*}
\end{equation}

\begin{figure}
\centering
\includegraphics[width=12.0cm]{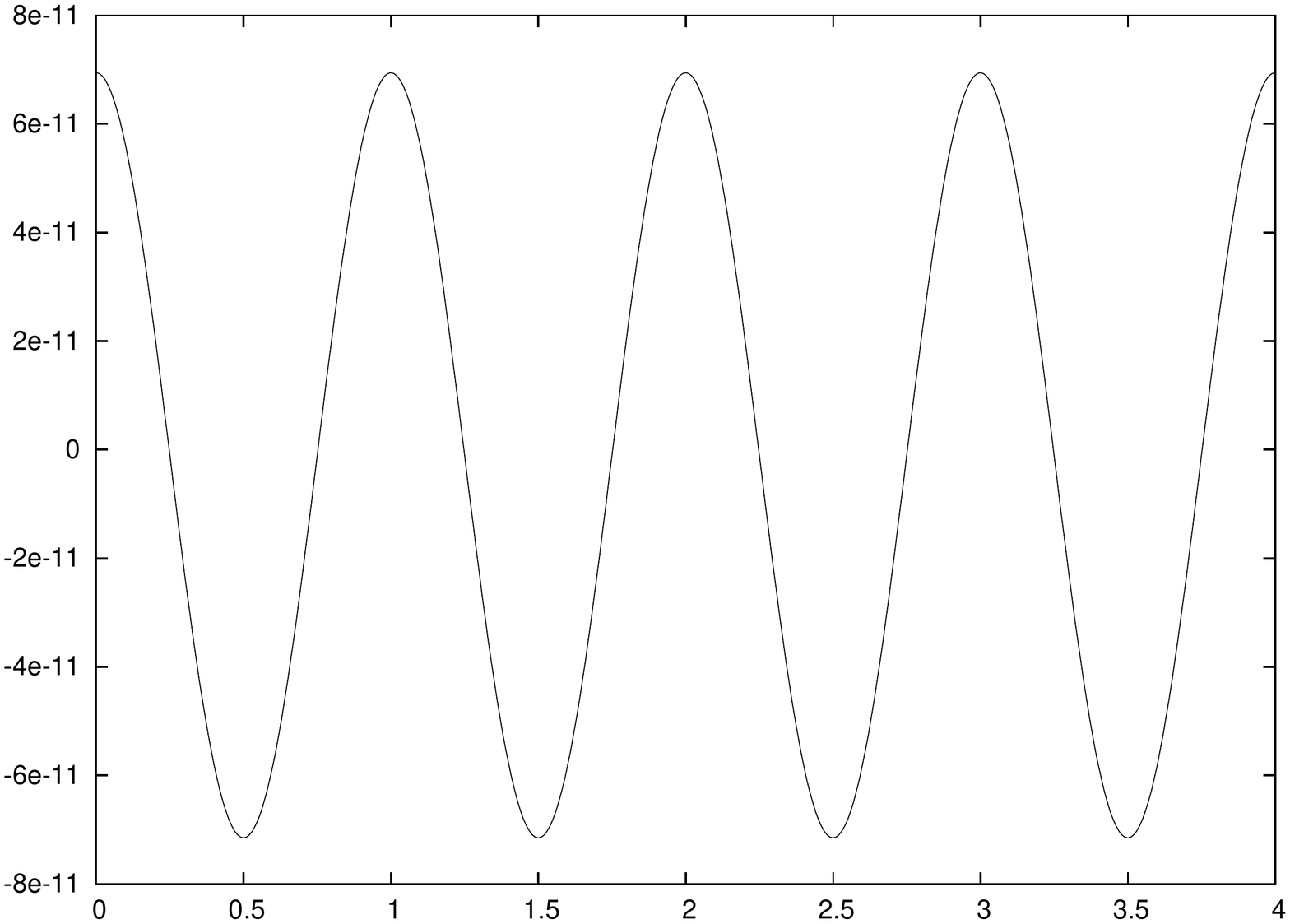}
\caption{Log-periodic oscillations in the~function $H^*(y)$ as a~function of $\log_2 y$, determined by numerically summing the~series in Equation (\ref{H*}), about the~mean value $1.44269504089$. Note the~small amplitude of the~oscillations. }
\label{figure3}
\end{figure}

The~function $H^*(y)$ is clearly a log-periodic function of $\log_2 (y)$ of period $1$.
A~plot of this function is shown in Figure~\ref{figure3}. We see that $H^*(y)$ is nearly a~constant, with value $a~\approx 1.
4426950409$, but it has small oscillations of amplitude of order $10^{-10}$. 

\vspace{-6pt}

We note that $H^*(y)$ remains bounded for all $y$, and hence $G(x=\frac{1}{1+y}) $ varies as $1/y^2$ for $y \rightarrow 0$. This implies that $T_n$ varies linearly with $n$ for large $n$. In addition, log-periodic oscillations of $T_n$ with $n$  correspond to~log-periodic oscillations of  $\tilde{H}(y)$ with $Y$. Write 
\begin{equation}
T_n \approx  a~~n \left[  1 + \alpha~\cos (2 \pi \log_2 n) \right]
\end{equation}

Then,
\begin{equation}
G(x) = a~x\frac{d}{dx}\left[ \sum_{n=0}^{\infty}  x^n (1 + \alpha~Re ~~ e^{ i 2 \pi \log_2 n}) \right]
\end{equation}

Using $\sum_n x^n n^z \sim (1 -x)^{-z-1}$, we see that this gives
\begin{eqnarray*}
 G(x) &\sim& \frac{a~}{(1-x)^2} + \alpha~Re \left[ \frac{1}{(1 -x)^{2 + \frac{2 \pi i}{\log 2}}}\right]\\
 & =& \frac{1 }{(1-x)^2} \left[ 1 + \alpha~\cos \log_2 (\frac{1}{1-x}) \right]
\end{eqnarray*}

 Define 
\begin{equation}
g(x)= \frac{\exp(x)}{[1+ \exp(x)]^2}.
\end{equation}

Then, we have 
\begin{equation}
H(y)= \sum_{n= -\infty}^{+\infty} g(n \log 2+\log y)
\end{equation}
 
Using  Poisson summation formula, we have
\begin{equation}
\sum_{n= -\infty}^{+\infty} g(nP +t) = \frac{1}{P} \sum_{k = -\infty}^{+\infty} \tilde{g}(k/P) \exp(2 \pi i k t/P)
\label{H_poisson}
\end{equation}
where $\tilde{g}(k)$ is the~Fourier transform of $g(x)$ defined as

\begin{equation}
\tilde{g}(k) = \int_{-\infty}^{+\infty} dx ~~ \frac{\exp(x)}{[1+ \exp(x)]^2} ~~\exp(2 \pi i k x) 
\label{FT}
\end{equation}

It is easy to~see that $\tilde{g}(0)=1$, which shows that $a~=1/\ln 2$. In~Equation (\ref{FT}), we note that the~integrand has simple poles in the~complex-x plane at $x = (2 n +1) \pi i$, for all integers $n$. For $n=0$, we may close the~contour from above. The~residue at the~poles $x = \pm i \pi $ is 
$-(1+\frac{2 \pi i}{ \ln 2})\exp(- \frac{2 \pi^2}{ \ln 2})$, which gives $\tilde{g}(1)= \frac{8\pi^2}{\ln 2}\exp(- \frac{2 \pi^2}{ \ln 2}) $, and  $\alpha~= a~ \tilde{g}(1) = 7.05 \times 10^{-11}$, which matches the~numerically observed value of amplitude of the~oscillations (Figure~\ref{figure2}).

\section{Summary and Concluding Remarks}
\label{section6}
In this paper, we have studied a~version of the~Minority game, where the~$(2N+1)$ agents try to~coordinate their actions to~get into~a~periodic state of period $(2N +1)$, in which every agent has the~same long-time average payoff, and the~global efficiency of the~system is the~maximum possible.
We have proposed an algorithm that can achieve this aim in a~time of order ${\cal O}(N)$. We were able to~determine the~average running time of this algorithm exactly. As the~time required to~coordinate is of order $N$, the~agents should have time horizon of at least order $N$, so that the~cost of reaching the~coordinated state is off-set by the~slightly higher payoffs later. 

As explained in the~Introduction,  we have assumed that the~ agents are rational, with unlimited memory, and use their full knowledge of efficiency of different strategies to~decide what to~do next. This is of course not realizable in practice. This work  only provides the~benchmark for measuring the~performance of imperfect agents.

The~question of how  all agents decide about which strategy they will all use is clearly important. If we {\it assume} that all agents will use the~strategy we have proposed, are we begging the~question of establishing coordination amongst them?  Our answer to~this question may not be not fully convincing to~all, and this issue perhaps needs further work. We have argued that the~strategy we have proposed is distinguished by its simplicity,  efficiency and analytical tractability. 

Let us consider an alternate algorithm  X that the~agents could use to~establish the~periodic state. 
For the~  purpose of describing this algorithm, let us further  assume that all agents have already somehow agreed that $B$ is always the~minority restaurant, and  that they should  use the~natural choice of periodic sequence $AABABABAB...AB$ of length $(2N+1)$,  with suitable phase shifts. The~aim is to~coordinate the~ choices of phase shifts  so that the~number of agents in B is exactly $N$ on all days.  However, now, the~agents try to~coordinate their  phase shifts by trial and error. Then, initially, each chooses a~phase shift  at random, from $0$ to~$2N$. They use their phase-shifted sequence  for $2N+1$~days, and, at the~end of this period,  take stock of the~attendance history of past $(2N+1)$~days. \linebreak If it is found that  the~restaurant $A$ had more than $N+1$ people on some day $r$. The~agent who was putting his phase shift starting at  Day $r$, with a~small probability, changes his phase shift to~another day,  randomly selected   out of the~ days that showed less crowding. 
Then, they watch the~performance over the~next $(2N+1)$ days, and again readjust phases again, until perfect coordination is achieved. It~is easy to~see that, while this will eventually find a~cyclic state, the~time required is much more than in our proposed~strategy.

The~problem of coordination amongst agents for optimal performance is also encountered in other games of resource allocation. For example, in the~Kolkata~Paise Restaurant problem {\color{black}\cite{kolkata, kolkata-book}}, one~has $N$ agents, and $N$ restaurants. The~agents are all equal, and all of them agree to~an agreed ranking of the~restaurants (i.e., $1$ to~$N$). It is also given that each restaurant only serves one customer per day at a~specially reduced price. Again, if the~agents cannot communicate directly with each other in choosing which restaurant to~go to, and they all want to~avail of the~special price, and also prefer to~go to~higher ranked restaurant, the~optimal state where each agent can get a~special price, and get to~sample higher ranked restaurants as often as others, is the~one where agents can organize themselves into~a~periodic state, where each agents visits restaurants ranked 1, 2, 3, and so on,  on successive days, in a~periodic way, and agents stagger the~phases of their cycles so that every restaurant has exactly one visitor on each day. In this case, the~problem of achieving the~cyclic state may also be reduced to~the~problem of assigning each of $N$ agents with a~unique ID between $1$ and~$N$. 
 
It is also straight forward to~extend this algorithm to~the~El Farol Bar problem,  where the~two states are Go to~bar, and Stay at home, and the~payoff is 1 to~agents who went to~the~bar, but only if the~attendance at the~bar is $\leq r$, where  $ r$ is not restricted to~be $N$.  Then, the~periodic state involving least number of switches is obtained when each agent goes to~bar for $r$ consecutive days, and stays at home for the~next $(2 N +1 -r)$ days, and the~agent  with ID $j$ phase-shifts the~origin of his cycle by amount $j$, $0 \leq j \leq 2N$.

Another point of interest in our results is the~finding of  log-periodic oscillations in the~average time. Log-periodic oscillations have been seen in the~analysis of several algorithms~\cite{log-periodic1, log-periodic2, odlyzko, log-periodic3}. In most of these cases, the~leading behavior is a~simple power-law (or logarithmic dependence), and the~log-periodic oscillations appear in the~first correction to~the~asymptotic behavior~\cite{footnote}. In the~problem studied here, the~oscillations are seen in the~coefficient of the~variation of the~leading linear dependence of average time of algorithm with number of agents.

 The~very small value of the~amplitude of oscillations deserves some comment. Firstly, normally, one would expect this to~be of $\mathcal{O}(1)$, and explaining the~origin of this  ``unnaturally small'' value  is  of some theoretical interest. Here, we could calculate this amplitude exactly, but  we do not know of any general argument to~estimate  even the~order of magnitude of such amplitudes,  without  doing the~exact~calculation.

In systems with discrete scale invariance, such as deterministic fractals, amplitudes of order ${\mathcal O}(10^{-2})$ have been seen~\cite{sumedha,akkermans}. However, even in these cases, where log-periodic oscillations are rather expected, they only form the~sub-leading correction: for example, in the~case of number rooted $P_n$ of polygons of length $n$ on fractals, we get $\log P_n$ has a~part that grows as $a~n + B \log n$, and then there is an additive log-periodic oscillatory term of finite amplitude~\cite{sumedha}. The~log-periodic oscillations in $T_n$ in this paper come {\it {in a~multiplicative term}}, and hence the~amplitude will grow with $n$, for large $n$. 
\vspace{6pt}

\section*{Author Contributions}{Deepak Dhar supervised the project and contributed towards the conceptualization of the proposed strategy, analysis and the overall writing, reviewing and editing process. Hardik Rajpal contributed towards the conceptualization, formal analysis, data visualization and the overall writing, reviewing and editing process.}

\acknowledgments{{Hardik Rajpal} thanks Tata~Institute of fundamental Research, Mumbai for hosting his visit in the~summer of 2016,  and Indian Institute of Science Education and Research, Pune, for support in the~summer of 2017, which made this work possible. We thank V. Sasidevan and R. Dandekar their comments on the~draft  manuscript. {Deepak Dhar}'s work is supported in part by the~J. C. Fellowship, awarded by the~Department of Science and Technology, India, under the~grant DST-SR-S2/JCB-24/2005.}


\bibliographystyle{unsrt}
\bibliography{Reference}

\end{document}